# Study the dynamics of the nonspreading Airy packets from the time evolution operator

Chyi-Lung Lin[1], Te-Chih Hsiung[2], Meng-Jie Huang[3]
[1]Department of Physics, Soochow University,
Taipei, Taiwan, R.O.C.
[2]Department of Physics, National Taiwan University,
Taipei, Taiwan, R.O.C.
[3]Magnetism Group / Scientific Research Division
National Synchrotron Radiation Research Center
Hsinchu, Taiwan, R.O.C.

## ABSTRACT

Berry and Balazs showed that an initial Airy packet $\mathbf{Ai}(\mathbf{b\,x})$ under time evolution is nonspreading in free space and also in a homogeneous time-varying linear potential $\mathbf{V}(\mathbf{x},\mathbf{t}) = -\mathbf{F}(\mathbf{t})\,\mathbf{x}$. We find both results can be derived from the time evolution operator U(t). We show that U(t) can be decomposed into ordered product of operators and is essentially a shift operator in x; hence, Airy packets evolve without distortion. By writing the Hamiltonian $\mathbf{H}$ as $\mathbf{H} = \mathbf{H_b} + \mathbf{H_i}$, where $\mathbf{H_b}$ is the Hamiltonian such that $\mathbf{Ai}(\mathbf{b\,x})$ is its eigenfunction. Then, $\mathbf{H_i}$ is shown to be as an interacting Hamiltonian that causes the Airy packet into an accelerated motion of which the acceleration $\mathbf{a} = (-\frac{\mathbf{H_i}}{\mathbf{x}})/\mathbf{m}$. Nonspreading Airy packet then acts as a classical particle of mass m, and the motion of it can be described classically by $\mathbf{H_i}$.

**Keywords**: time evolution; Airy function; nonspreading wave packet; Hamiltonian; Zassenhaus formula;

PACS numbers: 03.65. - w - Quantum mechanics
PACS numbers: 03.65. Ge - solution of wave equations: bound states

E-mail: cllin@scu.edu.tw

## I. Introduction

Airy functions are eigenfunctions of the Hamiltonian $\mathbf{H = H_0 + \ f\, x}$, where $\mathbf{H_0 = \frac{p^2}{2m}}$. Berry and Balazs showed the interesting phenomena of Airy functions under time evolution [1]. They showed that a wave $\boldsymbol{\Psi}$ in the form of an Airy function, then the propagation of $|\boldsymbol{\Psi}|^2$ is form invariant in two cases. One is in the free space and the other is in a potential $\mathbf{V(x,t) = -F(t)\, x}$. It had also been shown that Airy packets are nonspreading only in these two cases [2].

Why there are only these two cases? We may see this from the following consideration. If the Hamiltonian is simply as $(\mathbf{H_0 + f\, x})$, then Airy packets are stationary; they won't move. If the Hamiltonian is more than that, we then decompose the Hamiltonian into two parts. That is let $\mathbf{H = (H_0 + \ f\, x) + H_i}$. The first part doesn't drive Airy packets into motion. The second part does. For the free space, we have $\mathbf{H_i = -f\, x}$ . And for the case that $\mathbf{V(x,t) = -F(t)\, x}$, $\mathbf{H_i = -(f + F(t))\, x}$. For both cases, $\mathbf{H_i}$ is linear in x. One is time independent, and the other is time dependent. We don't have other cases of $\mathbf{H_i}$ that is linear in x.

We use this decomposed Hamiltonian to the time evolution operator, then we show that $\mathbf{H_i}$ which is linear in x, whether time independent or time dependent, does support nonspreading Airy packets. And we also show that the acceleration of Airy packets in these two cases can be obtained from $\mathbf{H_i}$.

The phenomenon that Airy packets propagate with a constant acceleration in free space may seem strange from the point of view of classical mechanics, and therefore seems to violate Ehrenfest's theorem in quantum mechanics [3]. However, this does not contradict Ehrenfest's theorem, as Airy packet is not square integrable. We therefore cannot use $|\boldsymbol{\Psi}|^2$ to describe the probability density for a single particle. As a result, the acceleration of the packet does not describe an accelerating particle.

The results of Berry and Balazs showed that if we start from an Airy packet at the time t = 0, that is, $\boldsymbol{\psi}\mathbf{(x,0) = Ai(b\, x)}$, then $\boldsymbol{\psi}\mathbf{(x,t)}$ is shown to be of the form

$$\boldsymbol{\psi}\mathbf{(x,t) = Ai[b\,(x - x_s(t))]e^{i\,\phi(x,t)}}. \qquad (1)$$

The initial packet $\mathbf{Ai(b\, x)}$ is now shifted to a packet $\mathbf{Ai[b\,(x - x_s(t))]}$ at time t. The term $\mathbf{x_s(t)}$ represents the amount of shift. Accompanied with this shift, there is also an extra phase factor $\mathbf{e^{i\,\phi(x,t)}}$ resulting from the time evolution.

We show in this paper that the form invariant and the acceleration of the Airy packets can be derived from the time evolution operator U(t). Formula (1) shows that the time evolution operator should effectively be of the form

$$\mathbf{U(t) = e^{i\,\phi(x,t)}\,exp\left[\frac{-i}{\hbar}\,x_s(t)\,p\right]}. \qquad (2)$$

We will show in sections III and IV that U(t) is indeed of the form as (2).

In Sec. II, we introduce some properties of Airy functions that are needed for later discussion. In Sec. III, we discuss the time evolution operator U(t) in free space. We use the fact that $\mathbf{Ai(b\,x)}$ is the eigenfunction of $\mathbf{H_b}$. Then we write $\mathbf{H_0 = H_b - f_b\,x}$ and using the Zassenhaus formula to decomposing U(t) into a product of operators. U(t) is then shown to contain essentially a phase factor and a shift operator. In Sec. IV, we discuss the case of U(t) in a potential $\mathbf{V(x,t) = -F(t)\,x}$. In this case, U(t) has already a known decomposed form. U(t) contains an extra shift operator other than the U(t) in free space. With the result derived in Sec. III, we easily obtain the results of Berry and Balazs. In Sec. V, we make a brief conclusion.

## II. Properties of Airy functions

Airy functions are defined by the following equation:

$$\mathbf{\frac{\partial^2}{\partial x^2}\,Ai(x) = x\,Ai(x)}. \qquad (3)$$

Above, we use the notation $\mathbf{Ai\,(x)}$ for the Airy function. For Airy function of a more general form $\mathbf{Ai[b\,(x - x_s)]}$, where $\mathbf{x_s}$ is some number or a function of time, the equation it satisfies is

$$\mathbf{\frac{\partial^2}{\partial x^2}\,Ai[b\,(x - x_s)] = b^3\,(x - x_s)\,Ai[b\,(x - x_s)]}. \qquad (4)$$

It is know that Airy function $\mathbf{Ai[b\,(x - x_s)]}$ is the eigenfunction of a Hamiltonian $\mathbf{H}$, and $\mathbf{H = H_0 + f_b\,x}$, where $\mathbf{f_b = b^3\hbar^2/(2m)}$ . We denote this Hamiltonian by $\mathbf{H_b}$. Then $\mathbf{Ai}[\mathbf{b}\ (\mathbf{x}-\mathbf{x_s})]$ is an eigenfunction of $\mathbf{H_b}$. The corresponding energy is $\mathbf{E_b} = \mathbf{f_b}\ \mathbf{x_s}$. We conclude all of this in the following formulas:

$$\mathbf{H_b} = \mathbf{H_0} + \mathbf{f_b}\ \mathbf{x}. \qquad (5)$$

$$\mathbf{H_b}\ \mathbf{Ai}[\mathbf{b}\ (\mathbf{x}-\mathbf{x_s})] = \mathbf{E_b}\ \mathbf{Ai}[\mathbf{b}\ (\mathbf{x}-\mathbf{x_s})]. \qquad (6)$$

$$\mathbf{f_b} = \frac{\mathbf{b^3}\hbar^2}{\mathbf{2m}}. \qquad (7)$$

$$\mathbf{E_b} = \mathbf{f_b}\ \mathbf{x_s}. \qquad (8)$$

## III. The time evolution in free space, $\mathbf{V(x,t) = 0}$

In free space U(t) is known to be

$$\mathbf{U(t)} = \mathbf{exp}\ \left[\frac{-\mathbf{i}}{\hbar}\mathbf{H_0 t}\right], \qquad (9)$$

$$\mathbf{H_0} = \frac{\mathbf{p^2}}{\mathbf{2m}} = -\frac{\hbar^2}{\mathbf{2m}}\frac{\partial^2}{\partial \mathbf{x^2}}\ .$$

We choose the initial wave function as $\boldsymbol{\psi}(\mathbf{x},\mathbf{0}) = \mathbf{Ai}(\mathbf{b\ x})$. Then $\boldsymbol{\psi}(\mathbf{x},\mathbf{t}) = \mathbf{U(t)}\ \mathbf{Ai}(\mathbf{b\ x})$. To calculate $\boldsymbol{\psi}(\mathbf{x},\mathbf{t})$, it would be complicate if we directly use U(t) in the form of (9). However, this can be simplified if we use $\mathbf{H_b}$ to operate on $\mathbf{Ai}(\mathbf{b\ x})$, as then we can use (6). We thus rewrite $\mathbf{H_0}$ as $\mathbf{H_0} = -\mathbf{f_b}\ \mathbf{x} + \mathbf{H_b}$. Then

$$\mathbf{U(t)} = \mathbf{exp}\ \left[\frac{-\mathbf{i}}{\hbar}(-\mathbf{f_b}\ \mathbf{x} + \mathbf{H_b}\ )\mathbf{t}\right]. \qquad (10)$$

We decompose (10) by using the Zassenhaus formula [4], which states that:

$$\mathbf{e^{A+B}} = \mathbf{e^A}\ \mathbf{e^B}\ \mathbf{exp}\left[-\frac{\mathbf{1}}{\mathbf{2}}[\mathbf{A},\mathbf{B}]\right]\mathbf{exp}\left[\frac{\mathbf{1}}{\mathbf{6}}[\mathbf{A},[\mathbf{A},\mathbf{B}]] + \frac{\mathbf{1}}{\mathbf{3}}[\mathbf{B},[\mathbf{A},\mathbf{B}]]\right]\ldots. \qquad (11)$$

Substituting $\mathbf{A} = \frac{\mathbf{i}}{\hbar}\ \mathbf{f_b}\ \mathbf{x}\ \mathbf{t}$, and $\mathbf{B} = \frac{-\mathbf{i}}{\hbar}\ \mathbf{H_b}\ \mathbf{t}$ to (10), U(t) is then decomposed into an ordered product of four terms. That is

$$U(t) = U(1)\, U(2)\, U(3)\, U(4)$$

$$\mathbf{U(1) = exp}\ \left[\frac{\mathbf{i}}{\hbar}\ \mathbf{f_b\ x\ t}\right]]. \qquad (12)$$

$$\mathbf{U(2) = exp}\left[\frac{\mathbf{-i}}{\hbar}\ \ \mathbf{H_b\ t}\right]. \qquad (13)$$

$$\mathbf{U(3) = exp}\left[\frac{\mathbf{-i}}{\hbar}\ \frac{\mathbf{f_b t^2}}{\mathbf{2m}}\ \mathbf{p}\right]. \qquad (14)$$

$$\mathbf{U(4) = exp}\left[\frac{\mathbf{i}}{\hbar}\ \frac{\mathbf{{f_b}^2\ t^3}}{\mathbf{6m}}\right]. \qquad (15)$$

We can easily read the result of the action of U(t) on $\mathbf{Ai\ (b\ x)}$. The operator U(4) gives a phase factor. The operator U(3) gives a shift in x, and then we have a shifted-packet $\mathbf{Ai\ [b\ (x - x_0(t))]}$, where

$$\mathbf{x_0(t) = \frac{f_b t^2}{2m}}. \qquad (16)$$

This packet $\mathbf{Ai\ [b\ (x - x_0(t))]}$ is an eigenfunction of $\mathbf{H_b}$, with energy $\mathbf{E_b = f_b\ x_0(t)}$. The following operator U(2) then gives a phase factor, and so thus U(1). We see that U(t) is essentially a shift operator. This explains why the Airy packet propagates without distortion.

We conclude that the action of U(t) on $\mathbf{Ai(b\ x)}$ is effectively as

$$\mathbf{U(t) = e^{i\ \phi_0(x,t)}\ exp}\ \left[\frac{\mathbf{-i}}{\hbar}\ \mathbf{x_0(t)\ p}\right] \qquad (17)$$

The phase $\mathbf{\phi_0(x, t)}$ can be calculated as

$$\mathbf{\phi_0(x,t)} = \frac{\mathbf{f_b\ t\ x}}{\hbar} + \frac{\mathbf{{f_b}^2 t^3}}{\mathbf{6m}\hbar} - \frac{\mathbf{E_b\ t}}{\hbar}$$

$$= \frac{\mathbf{f_b\ t}}{\hbar}\left(\mathbf{x} - \frac{\mathbf{f_b\ t^2}}{\mathbf{3m}}\right). \qquad (18)$$

We have the final result

$$\mathbf{\Psi(x,t) = exp}\ \left[\frac{\mathbf{-i}}{\hbar}\mathbf{H_0 t}\right]\ \mathbf{Ai\ (b\ x)}$$

$$= \mathbf{exp}[\mathbf{i}\ \boldsymbol{\phi}_{\mathbf{0}}(\mathbf{x},\mathbf{t})]\ \mathbf{Ai}\ [\mathbf{b}\ (\mathbf{x}-\mathbf{x}_{\mathbf{0}}(\mathbf{t}))]\ . \qquad (19)$$

where $\mathbf{x_0(t)}$ and $\boldsymbol{\phi}_\mathbf{0}(\mathbf{x},\mathbf{t})$ are those in (16) and (18). We will use (19) again in the next section.

From (19), the trajectory of the packet is described by $\mathbf{x} = \mathbf{x_0(t)} = \frac{\mathbf{f_b t^2}}{\mathbf{2m}}$, which represents a motion with an acceleration $\mathbf{a} = \frac{\mathbf{f_b}}{\mathbf{m}}$. To understand why the nonspreading packet gets an acceleration in free space, we note that the packet $\mathbf{Ai}\ (\mathbf{b\ x})$ is a stationary state when the Hamitonian is $\mathbf{H_b}$. That is, $\mathbf{H_b}$ contributes only a phase factor to Airy packets. In free space $\mathbf{H} = \mathbf{H_0}$. Writing $\mathbf{H_0} = \mathbf{H_b} + \mathbf{H_i}$, then we have $\mathbf{H_i} = -\mathbf{f_b\ x}$. Thus $\mathbf{H_0}$ contains besides $\mathbf{H_b}$ but also an extra term $\mathbf{H_i}$. As a result of that, the nonspreading quantum packet $\mathbf{Ai}\ (\mathbf{b\ x})$ should not be stationary in free space; it needs to move. It is this term $\mathbf{H_i}$ that causes the Airy packet into motion in free space, and into a motion with an acceleration. Thus to Airy packets, free space is in fact acting as a force field. The extra term $\mathbf{H_i}$ plays the role as an interacting Hamiltonian. A potential like $-\mathbf{f_b\ x}$ in classical mechanics offers a constant force, and therefore a constant acceleration. The motion of wave packets described in (16) may then have a classical-like interpretation. That is, we treat the nonspreading packet as a classical particle of mass m, then the force $\mathbf{f_b}$, which is resulted from the term $\mathbf{H_i} = -\mathbf{f_b\ x}$, will drive this particle into a motion with an acceleration $\mathbf{a} = \frac{\mathbf{f_b}}{\mathbf{m}}$. Thus, it is this term $\mathbf{H_i}$ that determines, in a classical way, the motion of the nonspreading wave packets. Such an interpretation can also be extended to the Sec IV.

The result in (19) is the same as that of Berry and Balazs, if we replace

$$\mathbf{b} \rightarrow \mathbf{B}\,\hbar^{-\frac{2}{3}}\ ,$$

$$\mathbf{f_b} = \frac{\mathbf{b^3}\hbar^2}{\mathbf{2m}} \rightarrow \frac{\mathbf{B^3}}{\mathbf{2m}}. \qquad (20)$$

## IV. The time evolution in a time-varying spatially uniform linear

**potential**

For the case $\mathbf{H(t) = H_0 + V(x,t)}$, where $\mathbf{V(x,t) = -F(t)\,x}$. The Schrodinger equation is

$$\mathbf{i\hbar \frac{\partial}{\partial t} \Psi(x,t) = H(t)\, \Psi(x,t).} \qquad (21)$$

$$\mathbf{H(t) = H_0 - F(t)\, x} \; . \qquad (22)$$

The differential equation for U(t) is

$$\mathbf{i\, \hbar \frac{\partial}{\partial t} U(t) = H(t)\, U(t).} \qquad (23)$$

with U(0) = 1. We cannot integrate (23) to give U(t) a simple form as that in (9). This is due to that $\mathbf{H(t)}$ is not commutative at two different times, as we have

$$\mathbf{[H(t_1), H(t_2)] = \frac{i\,\hbar}{m}\, p\, \big( F(t_2) - F(t_1) \big).} \qquad (24)$$

We then need a time-ordered product for the expression of U(t). However, following the suggestion of Baym [5], Gregorio and Castroto[6], showed that U(t) can be integrated , if we set

$$\mathbf{U(t) = e^{\frac{i}{\hbar}\alpha(t)\, x}\, \overline{U}}(t), \qquad (25)$$

$$\mathbf{\alpha(t) = \int_0^t F(\tau)\, d\tau.} \qquad (26)$$

The differential equation for $\mathbf{\overline{U}}$(t) is

$$\mathbf{i\, \hbar \frac{\partial}{\partial t} \overline{U}(t) = \frac{(p+\alpha(t))^2}{2m} \overline{U}(t).} \qquad (27)$$

We can now integrate (27) to get

$$\mathbf{\overline{U}(t) = \exp\left[\frac{-i}{\hbar} \int_0^t \frac{(p+\alpha(\tau))^2}{2m}\, d\tau\right]} \; . \qquad (28)$$

Substituting (28) to (25) and using $\frac{(\mathbf{p}+\boldsymbol{\alpha}(\boldsymbol{\tau}))^2}{\mathbf{2m}} = \frac{\boldsymbol{\alpha}(\boldsymbol{\tau})^2}{\mathbf{2m}} + \frac{\boldsymbol{\alpha}(\boldsymbol{\tau})}{\mathbf{m}}\mathbf{p} + \mathbf{H_0}$, we can again decompose U(t) into an ordered product of four terms, that is

U(t) = U(1) U(2) U(3) U(4),

$$\mathbf{U(1) = exp}\left[\frac{\mathbf{i}}{\hbar}\ \boldsymbol{\alpha}(\mathbf{t})\ \mathbf{x}\right]. \qquad (29)$$

$$\mathbf{U(2) = exp}\left[\frac{-\mathbf{i}}{\hbar}\int_{\mathbf{0}}^{\mathbf{t}}\frac{\boldsymbol{\alpha}(\boldsymbol{\tau})^2}{\mathbf{2m}}\mathbf{d}\boldsymbol{\tau}\right]. \qquad (30)$$

$$\mathbf{U(3) = exp}\left[\frac{-\mathbf{i}}{\hbar}\int_{\mathbf{0}}^{\mathbf{t}}\frac{\boldsymbol{\alpha}(\boldsymbol{\tau})}{\mathbf{m}}\ \mathbf{d}\boldsymbol{\tau}\ \mathbf{p}\right]. \qquad (31)$$

$$\mathbf{U(4) = exp}\left[\frac{-\mathbf{i}}{\hbar}\ \mathbf{H_0}\ \mathbf{t}\right] \qquad (32)$$

We see U(t) essentially contains only an extra shift operator U(3) relative to that in free space . We can read the result of the action of U(t) on the initial packet $\mathbf{Ai[b\ x]}$. The action of U(4) gives the result in (19). Thus

$$\mathbf{U(4)\ Ai[b\ x] = Ai\ [b\ (x - x_0(t))]\ exp\ [i\ \boldsymbol{\phi}_0(x,t)]}. \qquad (33)$$

The following action of U(3) will introduce a shift in x by an amount $\mathbf{x_1(t)}$ to both terms in (33); as both contain x. From (31), we have

$$\mathbf{x_1(t)} = \int_{\mathbf{0}}^{\mathbf{t}}\frac{\boldsymbol{\alpha}(\boldsymbol{\tau})}{\mathbf{m}}\ \mathbf{d}\boldsymbol{\tau} = \int_{\mathbf{0}}^{\mathbf{t}}\int_{\mathbf{0}}^{\boldsymbol{\tau}}\frac{\mathbf{F(s)}}{\mathbf{m}}\mathbf{ds\ d}\boldsymbol{\tau}. \qquad (34)$$

From (34), we see that $\mathbf{x_1(t)}$ represents a motion with an acceleration $\mathbf{a} = \frac{\mathbf{F(t)}}{\mathbf{m}}$. The other form of $\mathbf{x_1(t)}$ is

$$\mathbf{x_1(t)} = \int_{\mathbf{0}}^{\mathbf{t}}\frac{\mathbf{F}(\boldsymbol{\tau})}{\mathbf{m}}(\mathbf{t} - \boldsymbol{\tau})\mathbf{d}\boldsymbol{\tau}, \qquad \mathbf{(35)}$$

which is the form used in Berry and Balazs's paper. Using (18), we have therefore

$$\mathbf{U(3)U(4)\ Ai[b\ x]}$$

$$= \mathbf{Ai}\,[\mathbf{b}\,(\mathbf{x}-\mathbf{x_0}(\mathbf{t})-\mathbf{x_1}(\mathbf{t}))]\,\mathbf{exp}\,\left[\mathbf{i}\,\boldsymbol{\phi}_\mathbf{0}(\mathbf{x},\mathbf{t})-\frac{\mathbf{f_b\,t}}{\hbar}\mathbf{x_1}(\mathbf{t})\right]. \qquad (36)$$

The overall action of U(t)) then gives the final result:

$$\mathbf{U}(\mathbf{t})\,\mathbf{Ai}[\mathbf{b\,x}]$$
$$= \mathbf{Ai}\,[\mathbf{b}\,(\mathbf{x}-\mathbf{x_0}(\mathbf{t})-\mathbf{x_1}(\mathbf{t}))]\ \ \mathbf{exp}[\mathbf{i}\,\boldsymbol{\phi}(\mathbf{x},\mathbf{t})]. \qquad (37)$$

The total phase $\boldsymbol{\phi}(\mathbf{x},\mathbf{t})$ is from the phases in U(1) U(2) and (36), we have

$$\boldsymbol{\phi}(\mathbf{x},\mathbf{t}) = \boldsymbol{\phi}_\mathbf{0}(\mathbf{x},\mathbf{t})-\frac{\mathbf{f_b\,t}}{\hbar}\mathbf{x_1}(\mathbf{t})+\frac{\mathbf{1}}{\hbar}\,\boldsymbol{\alpha}(\mathbf{t})\,\mathbf{x}-\frac{\mathbf{1}}{\hbar}\int_\mathbf{0}^\mathbf{t}\frac{\boldsymbol{\alpha}(\boldsymbol{\tau})^\mathbf{2}}{\mathbf{2m}}\,\mathbf{d}\boldsymbol{\tau}. \qquad (38)$$

Our final result is described in (37-38). From (37), we see that the packet $\mathbf{Ai}[\mathbf{b\,x}]$ under time evolution is moving in a trajectory described by $\mathbf{x}(\mathbf{t}) = \mathbf{x_0}(\mathbf{t})+\mathbf{x_1}(\mathbf{t})$. This is a motion with an acceleration $\mathbf{a} = \frac{\mathbf{f_b}}{\mathbf{m}}+\frac{\mathbf{F}(\mathbf{t})}{\mathbf{m}}$. To understand how nonspreading packet gets this acceleration from $\mathbf{H}(\mathbf{t})$, we rewrite $\mathbf{H}(\mathbf{t})$ in (22) as $\mathbf{H}(\mathbf{t}) = \mathbf{H_b}+\mathbf{H_i}$, where $\mathbf{H_i} = -(\mathbf{f_b}+\mathbf{F}(\mathbf{t}))\,\mathbf{x}$. In this form, the quantum packet will effectively be driven into accelerated motion by the term $\mathbf{H_i}$. We may similarly have a classical-like interpretation as that in Section III. We may treat the nonspreading packet as a classical particle of mass m, then the force $\mathbf{f_b}+\mathbf{F}(\mathbf{t})$, which is resulted from $\mathbf{H_i} = -(\mathbf{f_b}+\mathbf{F}(\mathbf{t}))\,\mathbf{x}$, will drive this classical-like particle into a motion with an acceleration $\mathbf{a} = (\mathbf{f_b}+\mathbf{F}(\mathbf{t}))/\mathbf{m}$.

Using (20) then shows that our result is equivalent to that of Berry and Balazs.

## V. Conclusion

The motion of a nonspreading wave packet should be similar to a classical particle, as it does not change its shape during the motion. Is it that the dynamics of the nonspreading packet the same as that for a classical particle?

We have shown above how to properly interpreting the Hamitonian $\mathbf{H}(\mathbf{t})$. For $\mathbf{H}(\mathbf{t}) = \mathbf{H_0}+\mathbf{V}(\mathbf{x},\mathbf{t})$, in general, we treat $\mathbf{V}(\mathbf{x},\mathbf{t})$ as the interacting Hamitonian. However, when discussing nonspreading Airy packets, we should decompose $\mathbf{H(t)}$ as $\mathbf{H(t) = H_b + H_i}$, then $\mathbf{H_i}$ plays the role as interacting Hamitonian. Thus for free space, though $\mathbf{V(x,t) = 0}$, yet, $\mathbf{H_i \neq 0}$, and therefore Airy packet accelerates. For $\mathbf{V(x,t) = -F(t)\,x}$, then $\mathbf{H_i = -(f_b + F(t))\,x}$. Thus Airy packet accelerates; however, the acceleration is not of the value $\mathbf{a = F(t)/m}$, but of the value $\mathbf{a = (f_b + F(t))/m}$. It is $\mathbf{H_i}$ not $\mathbf{V(x,t)}$ that determines the acceleration of nonspreading packet. We conclude that a nonspreading Airy packet acts as classical particle of mass m, and its motion can be described classically by $\mathbf{H_i}$.